\documentclass[aps,prb,reprint,longbibliography]{revtex4-2}
\usepackage[english]{babel}
\usepackage{upgreek}
\usepackage{amssymb}
\usepackage{amsmath,amsfonts,amsthm,amsbsy,empheq}
\usepackage{mathtools,bm}
\usepackage[caption=false]{subfig}
\usepackage{graphicx,bm}
\usepackage{physics}
\usepackage{dsfont}
\usepackage{bibentry}
\usepackage{xcolor}
\usepackage{breqn}
\usepackage{ulem}
\usepackage{standalone}
\usepackage{tikz}
\usepackage{grffile}
\usetikzlibrary{calc,decorations.markings,arrows.meta,shadings,patterns}

\DeclareTextCompositeCommand{\k}{LY1}{e}
{\oalign{e\crcr\noalign{\kern-.27ex}\hidewidth\char7\hidewidth}}

\usepackage[colorlinks,linkcolor=red]{hyperref}
\usepackage{ulem}
\usepackage{orcidlink}

\newif\ifrevision
\revisiontrue   

\ifrevision
  
  \newcommand{\del}[1]{\textcolor{red}{\sout{#1}}}
  
  \newcommand{\rnote}[1]{\textcolor{magenta}{\textbf{[#1]}}}
\else
  
  \newcommand{\del}[1]{}
  
  \newcommand{\rnote}[1]{}
  
\fi

\def\tr{\ensuremath{\operatorname{tr}}}

\makeatletter
\let\cat@comma@active\@empty
\makeatother
\begin{document}
	
	\title{Engineering symmetry-protected  topological states in waveguide arrays}
	
	\author{Lavi K. Upreti\,\orcidlink{0000-0002-1722-484X}}
    \email[]{lavi.upreti@uzh.ch}
	\affiliation{Department of Physics, University of Z\"urich, 8057 Z\"urich, Switzerland}

	
	\date{\today}	
    \begin{abstract}
    The topological classification of a system depends on the discrete symmetries of its Hamiltonian. In Floquet photonic waveguide arrays, the abstract symmetries of the Altland-Zirnbauer (AZ) scheme—chiral, particle-hole, and time-reversal (for photonics, $z$-reversal)—arise from structural properties of the lattice, yet a systematic correspondence has not been established. Here, we illustrate this correspondence for a simpler system of one-dimensional waveguide arrays with real coupling coefficients, showing how bipartite structure and $z$-reflection symmetry alone determine the whole AZ class. We further demonstrate that non-bipartite networks—lacking conventional particle-hole symmetry, chiral symmetry, and $z$-reversal symmetry—can nonetheless support topologically protected boundary states at quasienergy $\varepsilon = \pi$, even in one-dimension. The protecting symmetry—\textit{shifted}-particle-hole symmetry applies equally to higher-dimensional Floquet waveguides.
\end{abstract}
	\maketitle

    \section{Introduction}\label{sec:intro}

    Topological phases of matter support boundary states that are robust against local perturbations, protected by invariants that depend on the symmetries of the Hamiltonian~\cite{Hasan2010,Qi2011}. The AZ tenfold classification organizes these single-particle Hamiltonians according to time-reversal ($\mathcal{T}$), particle-hole ($\mathcal{C}$), and chiral ($\Gamma = \mathcal{T} \cdot \mathcal{C}$) symmetries, determining which topological invariants can be nontrivial in each class~\cite{Altland1997,Schnyder2008,Kitaev2009,Ryu2010,Chiu2016}. Crystalline symmetries further enrich this classification~\cite{Fu2011,Chiu2013,Shiozaki2014}.
    
    In photonic waveguide arrays, the propagation coordinate $z$ serves as an effective time, and periodic $z$-modulation implements Floquet driving~\cite{Garanovich2012,Rechtsman2013,Lu2014,Ozawa2019}. Experiments have demonstrated photonic Floquet topological insulators~\cite{Rechtsman2013}, anomalous edge states~\cite{Maczewsky2017}, and topological phenomena in non-linear regimes~\cite{Mukherjee2020,Mukherjee2021,Jurgensen2021,Kirsch2021,Szameit2024}. The abstract symmetries of the AZ classification, however, originate in fermionic electron-hole duality and do not directly apply to bosonic photons. Although partial structural interpretations exist in the literature, with bipartite lattices implying chiral symmetry\cite{Asboth2016} and $z$-reflection combined with real couplings implying effective $z$-reversal symmetry~\cite{Bellec2017}, a systematic correspondence between crystalline structure and AZ classes in photonic Floquet systems has not yet been established.

    Several extensions of topological protection exploit non-symmorphic symmetries. Time-glide symmetry—spatial reflection composed with half-period translation—protects Floquet phases absent in static systems~\cite{Morimoto2017}. Generalized chiral symmetry applies to unitcells with more than two sites~\cite{Ni2019}, and boundary states can survive when symmetry holds only on a sublattice~\cite{Chen2023}. None of these works, however, considers a variant of AZ symmetry with a momentum shift—a possibility that arises naturally in non-bipartite lattices.
    
    In this paper, we present two main results. In section~\ref{sec:symph}, we establish the systematic correspondence between structural lattice properties and AZ symmetries for one-dimensional photonic waveguide arrays with real couplings. In the same section, we also define \textit{shifted} particle-hole symmetry and show that it protects $\varepsilon = \pi$ boundary states in non-bipartite networks where conventional AZ symmetries are absent. Section~\ref{sec:engsymph} demonstrates this symmetry in a three-waveguide network, and we conclude in section~\ref{sec:conc}.

	\section{Symmetries in photonic waveguide array}\label{sec:symph}
    
\begin{figure}[!htb]
                \includegraphics[scale=0.46]{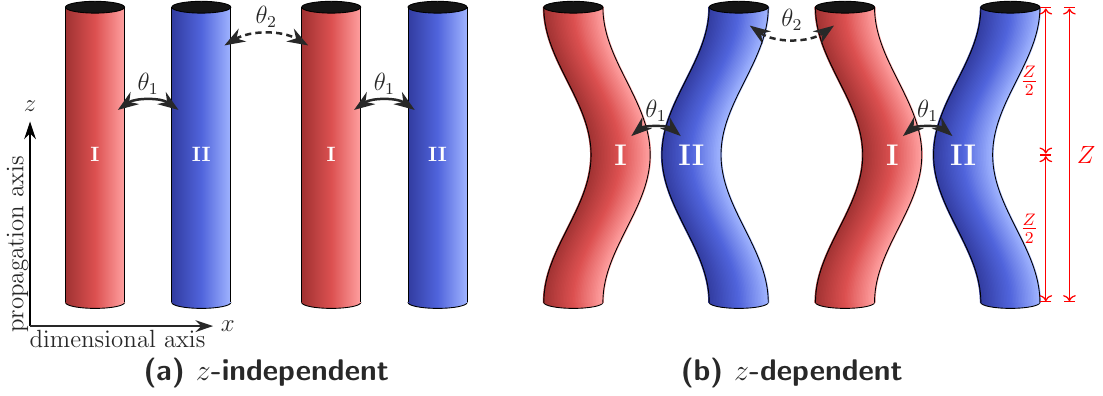}
		\caption{One-dimensional waveguide array with (a) $z$-independent coupling, realizing a static effective Hamiltonian, and (b) $z$-periodic coupling with period $Z$, realizing a Floquet-type driven system.}
		\label{fig:photonic}
	\end{figure}

    We consider one-dimensional optical waveguide arrays with evanescent coupling, where the refractive index is modulated periodically along one spatial direction (the $x$-axis) while light propagates along the $z$-axis. In this geometry, the paraxial wave equation assumes the form of a Schrödinger equation with the propagation coordinate $z$ playing the role of time~\cite{Longhi2009,Szameit2010}. This becomes clear when we consider envelope $\psi(\textbf{r}_\perp,z)$ that varies on length scales much greater than wavelength $\lambda$ of the optical field $E(\textbf{r}_\perp,z)$. Then the optical field is decomposed to $\psi(\textbf{r}_\perp,z)\,e^{i k_0 n_0 z}$, where $\psi(\textbf{r}_\perp,z)$ obeys
        \begin{equation}\label{eq:schrod}
            i\lambdabar \frac{\partial \psi}{\partial z} 
            = -\frac{\lambdabar^2}{2 n_0}\nabla_\perp^2 \psi 
            - \Delta n(\textbf{r}_\perp,z)\,\psi,
        \end{equation}
    where $\textbf{r}_\perp = (x,y)$ is the transverse coordinate, $\lambdabar = \lambda/(2\pi)$ is the reduced wavelength, $n_0$ is the  bulk refractive index, and $\Delta n(\textbf{r}_\perp,z)$ is the local 
    refractive index contrast defining the waveguides. The mapping to a  Schr\"odinger equation identifies $\lambdabar^2/(2 n_0)$ with  $\hbar^2/(2m)$ and $-\Delta n$ with the confining potential. Three 
    approximations underlie this description: the field is treated as scalar, neglecting polarization mixing, the longitudinal second derivative $\partial_z^2 \psi$ is dropped, and each waveguide supports a single transverse mode. The array thus realizes a tight-binding lattice with one orbital per site~\cite{Rechtsman2013}. In this lattice, the on-site energy is controlled by the local refractive index. The hopping amplitude comes from the evanescent overlap of neighboring modes and decays exponentially with the inter-waveguide separation.
     To see this, when the coupling strength $\theta$ remains uniform along the propagation axis—as illustrated in Fig.~\ref{fig:photonic}—the effective Hamiltonian is $z$-independent, analogous to a static quantum system. Conversely, when the coupling varies with $z$, the Hamiltonian becomes $z$-dependent. We consider periodic modulation, as shown in Fig.~\ref{fig:photonic}, in which the coupling parameters $\theta_{1,2}$ repeat with period $Z$, realizing a photonic analog of the Floquet version~\cite{Rechtsman2013,Rudner2020}.
%

%

    In both cases, the topological classification follows from three fundamental symmetries of the effective Hamiltonian: time-reversal, chiral, and particle-hole symmetry. In photonic waveguide arrays, the time-reversal symmetry of quantum mechanics is replaced by \textit{z}-reversal symmetry ($z$-RS), denoted by the operator $\mathcal{T}_z$, which reverses the propagation direction.\footnote{In lossless dielectric systems, Maxwell's equations are invariant under physical time reversal. This symmetry is therefore always present and cannot distinguish topological phases. The relevant symmetry for classification in paraxial systems is instead reversal of the propagation coordinate $z$, which plays the role of time in the effective Schrödinger equation~\cite{Rechtsman2013}.} For $z$-dependent Hamiltonians, $\mathcal{T}_z$, chiral symmetry $\Gamma$ and particle-hole symmetry $\mathcal{C}$ constraints the $z$-dependent Bloch Hamiltonian $H(k,z)$—where $k$ is the crystal momentum corresponding to the periodic lattice direction along $x$—as follows~\cite{Ryu2010,Chiu2016}:
    \begin{eqnarray}\label{eq:symm}
            T_z\,H(k,z) \, T_z^{-1} &=& H(-k,-z), \nonumber\\
            \Gamma\, H(k,z)\, \Gamma^{-1} &=& -H(k,-z),\nonumber \\
            C\, H(k,z)\, C^{-1} &=& -H(-k,z),
        \end{eqnarray}
        where $T_z = \mathcal{T}_z\,\mathcal{K}$ and $C = \mathcal{C}\,\mathcal{K}$ are 
        anti-unitary, with $\mathcal{K}$ complex conjugation and $\mathcal{T}_z$, 
        $\mathcal{C}$ the unitary parts. In terms of the unitary parts the 
        conditions read
        \begin{eqnarray}\label{eq:symm-unitary}
            \mathcal{T}_z\, H(k,z)\, \mathcal{T}_z^{-1} &=& H^{*}(-k,-z), \nonumber\\
            \Gamma\, H(k,z)\, \Gamma^{-1} &=& -H(k,-z),\nonumber \\
            \mathcal{C}\, H(k,z)\, \mathcal{C}^{-1} &=& -H^{*}(-k,z).
        \end{eqnarray}
    These symmetry constraints determine the AZ class of the system and, consequently, which topological invariants may be nontrivial~\cite{Schnyder2008,Kitaev2009}. The constraints themselves are linked to the structural properties of the waveguide lattice. We now introduce two such properties: bipartite lattice structure and $z$-reflection symmetry.




    \subsection{Structural properties}\label{sec:crystalline}
    
        \subsubsection{Bipartite structure (BpS)}\label{sec:bps}

\begin{figure}[!htb]
            \centering
            \includegraphics[scale=0.5]{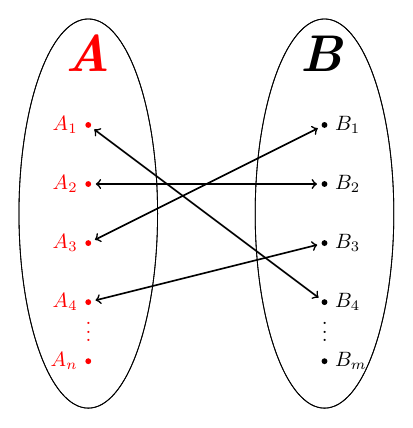}
            
            \caption{Bipartite lattice structure: coupling exists only between the $A$ and $B$ families, with no intra-family connections.}
            \label{fig:bipfig}
        \end{figure}

        A lattice possesses sublattice or bipartite structure when its sites can be partitioned into two disjoint families, labeled $A$ and $B$, such that coupling occurs only between families—never within the same family. This structure is illustrated in Fig.~\ref{fig:bipfig}. Canonical examples include the Su-Schrieffer-Heeger (SSH) chain in one dimension~\cite{Su1980,Asboth2016} and, in two dimensions, the honeycomb~\cite{Kane2005} and Lieb~\cite{Weeks2010} lattices. 

        To illustrate BpS, let us denote the orthogonal projectors onto the two sublattices by $P_A$ and $P_B$ (with $P_A + P_B = \mathds{1}$). BpS requires that the Hamiltonian satisfies
        \begin{equation}\label{eq:bpscs}
            P_A H P_A = P_B H P_B = 0,
        \end{equation}
        so that $H = P_A H P_B + P_B H P_A$. All matrix elements of $H$ reside in the off-diagonal blocks when expressed in the sublattice basis.

        For a unit cell containing $n$ sites of sublattice $A$ and $m$ sites of sublattice $B$, the Hamiltonian takes the general block off-diagonal form
        \begin{equation}\label{eq:H_chiral}
            H = \begin{pmatrix}
                0_{n\times n} & D \\
                D^\dagger & 0_{m\times m}
            \end{pmatrix},
        \end{equation}
        where $D$ is an $n \times m$ matrix encoding the inter-sublattice couplings. This structure implies the symmetry relation
        \begin{equation}\label{eq:bps_ham}
            \Sigma_z H \Sigma_z^{-1} = -H,
        \end{equation}
        with the sublattice operator
        \begin{equation}\label{eq:Sigma_z}
            \Sigma_z = \begin{pmatrix}
                \mathds{1}_{n\times n} & 0 \\
                0 & -\mathds{1}_{m\times m}
            \end{pmatrix}.
        \end{equation}

        The anticommutation relation $\{\Sigma_z, H\} = 0$ is the key defining property of chiral symmetry in the AZ classification~\cite{Ryu2010}; the sublattice operator $\Sigma_z$ therefore serves as the chiral symmetry operator $\Gamma$ introduced in Sec.~\ref{sec:symph}. BpS alone places the system in class AIII, characterized by a $\mathbb{Z}$ winding number. Additional crystalline symmetries further enrich the topological structure. We now consider one such symmetry.

    \subsubsection{\textit{z}-Reflection symmetry ($z-$Ref)}
    
    We now turn to the second structural property that is the $z$-reflection symmetry. A waveguide array possesses $z-$Ref symmetry if there exists an axis $z = z_0$ about which the couplings are mirror-symmetric along $z$: the system at $z_0 + \delta z$ is identical to that at $z_0 - \delta z$. This symmetry, represented by a unitary operator $\mathcal{R}_z$ that acts on the spatial configuration of the array, constrains the Hamiltonian according to
    \begin{equation}\label{eq:zref}
        \mathcal{R}_z H(k, z_0 + z) \mathcal{R}_z^{-1} = H(k, z_0 - z).
    \end{equation}
    When the reflection axis coincides with the origin ($z_0 = 0$), this simplifies to
    \begin{equation}\label{eq:zref_origin}
        \mathcal{R}_z H(k, z) \mathcal{R}_z^{-1} = H(k, -z).
    \end{equation}
    
\begin{figure}[!htb]
        \centering
        \includegraphics[scale=0.4]{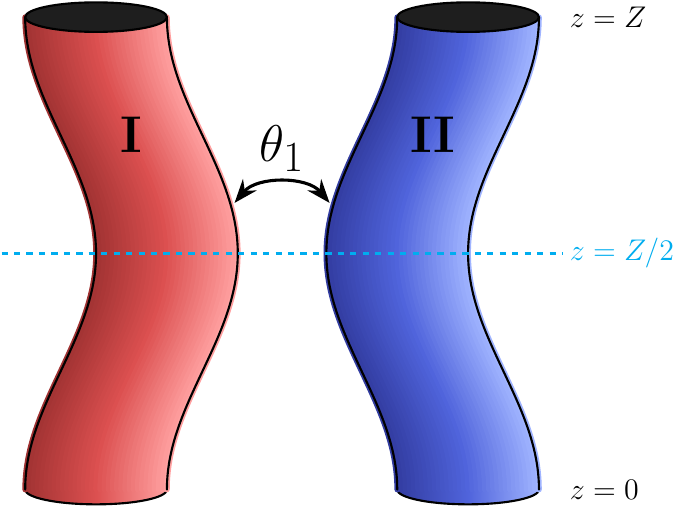}
        \caption{Reflection symmetry points within one period: the coupling configuration is symmetric about both $z = 0$ and $z = Z/2$.}
        \label{fig:z_ref_unitcell}
    \end{figure}

    For the periodically modulated array in Fig.~\ref{fig:photonic}\bm{$\mathsf{(b)}$}, $z$-Ref symmetry holds at $z_0 = 0$ and $z_0 = Z/2$, as illustrated in Fig.~\ref{fig:z_ref_unitcell}.
    
    The interplay between $z-$Ref symmetry and BpS determines the full symmetry class, as we establish below. When the coupling parameters are real—as is typical for evanescent coupling between dielectric waveguides—$z$-Ref symmetry implies $z$-reversal symmetry: complex conjugation leaves the Hamiltonian invariant, so that $\mathcal{R}_z H(k,z) \mathcal{R}_z^{-1} = H(k,-z)$ becomes $\mathcal{T}_z H(k,z) \mathcal{T}_z^{-1} = H^*(-k,-z)$ with $\mathcal{T}_z = \mathcal{R}_z \mathcal{K}$, where $\mathcal{K}$ denotes complex conjugation. The system then possesses all three symmetries of the BDI class: $z$-reversal, chiral, and particle-hole symmetry, which supports a $\mathbb{Z}$ topological invariant in one dimension~\cite{Schnyder2008,Chiu2016}.

    \subsection{Discrete symmetries in photonic systems}\label{subsec:Discrete}
    The structural properties introduced above—BpS and $z$-Ref symmetry—give rise to discrete symmetries that determine the topological classification. In this section, we formalize the three fundamental symmetries of the AZ scheme—chiral, $z$-reversal, and particle-hole symmetry—and establish their relationship to the structural properties of waveguide arrays.
        
        \subsubsection{Chiral symmetry (CS)}
        
        For a $z$-independent Hamiltonian $H$, CS is defined by a unitary operator $\Gamma$ satisfying the anticommutation relation
        \begin{equation}\label{eq:css}
            \{H, \Gamma\} = 0 \quad \Longleftrightarrow \quad \Gamma H \Gamma = -H,
        \end{equation}
        where $\Gamma = \Gamma^\dagger = \Gamma^{-1}$, implying $\Gamma^2 = \mathds{1}$.\footnote{This phase convention can always be achieved: if initially $\Gamma^2 = e^{i\phi}$, redefine $\Gamma \to e^{-i\phi/2}\Gamma$.} As established in Sec.~\ref{sec:bps}, this condition is automatically satisfied by the sublattice operator $\Sigma_z$ when the Hamiltonian possesses BpS. In the sublattice basis, we may identify
        \begin{equation}\label{eq:cbs}
        \Gamma = P_A - P_B = \Sigma_z,
        \end{equation}
        so that bipartite lattice geometry with purely inter-sublattice hopping implies CS~\cite{Asboth2016,Ryu2010}. The spectral constraint is that eigenvalues appear in $\pm\mathcal{E}$ pairs: if $|\psi\rangle$ is an eigenstate with energy $\mathcal{E}$, then $\Gamma|\psi\rangle$ is an eigenstate with energy $-\mathcal{E}$. At zero energy, eigenstates can be sublattice-polarized—residing entirely on one sublattice family—which underlies the spatial localization of topological boundary modes~\cite{Su1980,Asboth2016}.



        \textit{Extension to $z$-dependent systems.}
        For $z$-dependent Hamiltonians, CS about the origin $z_0 = 0$ takes the form
        \begin{equation}\label{eq:csd}
        \Gamma H(k,z) \Gamma^{-1} = -H(k,-z),
        \end{equation}
        where the sign flip in the $z$-argument reflects the non-local character of CS along the propagation direction. 
        
        The corresponding constraint on the evolution operator $U(k,z)$ reads
        \begin{equation}
            \Gamma U(k,z) \Gamma^{-1} = U(k,-z).
        \end{equation}
        This follows from the definition $U(k,z) = \mathcal{T} \exp\bigl[-i\int_0^z dz' H(k,z')\,\bigr]$ and the anticommutation of $\Gamma$ with $H$.
        
        For a periodic Hamiltonian with period $Z$ in $z$, this condition evaluated about a symmetry point $z_0$ becomes
        \begin{equation}\label{eq:G_chiral_z}
            \Gamma H(k, z_0 + z) \Gamma^{-1} = -H(k, Z + z_0 - z).
        \end{equation}

        Eq.~\eqref{eq:G_chiral_z} identifies \textit{chiral-symmetric points} $z_0$ where the symmetry acts locally. For a $Z$-periodic system, these occur at $z_0 = 0$ and $z_0 = Z/2$, where the evolution operator satisfies 
        \begin{equation}
            \Gamma\, U(z - z_0, -z_0)\, \Gamma^{-1} = U^\dagger(z + z_0, z_0).
        \end{equation}
        Setting $z = Z$ yields the Floquet operator constraint $\Gamma\, U_F\, \Gamma^{-1} = U_F^{\dagger}$, which ensures quasienergy pairing $\varepsilon \leftrightarrow -\varepsilon$. 
        Identifying such symmetric points may require a judicious choice of origin; for instance, if $H = H_0 + V\sin(\omega z)$, shifting $z \to z + \pi/(2\omega)$ transforms the driving to $V\cos(\omega z)$, which is symmetric about the origin $z = 0$.

        \textit{Decomposition into structural symmetries.}
        In Floquet systems, CS arises from two ingredients: BpS ($\Sigma_z$) (cf. Eq.\eqref{eq:cbs}) and $z$-Ref symmetry. The combined action yields
        \begin{equation}\label{eq:H_chiral_z}
        \Sigma_z \mathcal{R}_z H(k, z_0 + z) \mathcal{R}_z^{-1} \Sigma_z^{-1} = -H(k, z_0 - z).
        \end{equation}
        The chiral operator about the symmetry point $z_0$ is thus $\Gamma = \Sigma_z \mathcal{R}_z$, combining the sublattice and time reflection operations. This composite structure—distinct from the purely sublattice-based chiral operator of static systems—reflects the interplay between BpS and $z$-Ref in Floquet systems.

        \textit{CS without BpS and $z$-Ref.}
        Remarkably, CS in $z$-dependent case can persist even when both BpS and $z$-Ref symmetry are simultaneously broken. Consider the Hamiltonian
        \begin{equation}\label{eq:nbcs}
        H(k,z) = \begin{pmatrix}
        V\sin(z) & J_1\cos(z) + J_2 e^{ik} \\[1mm]
        J_1\cos(z) + J_2 e^{-ik} & -V\sin(z)
        \end{pmatrix}.
        \end{equation}
        The diagonal potential $V\sin(z)$ breaks BpS (due to nonzero diagonal elements), while the particular $z$-dependence of both diagonal and off-diagonal terms breaks $z$-Ref symmetry. 
        Nevertheless, this Hamiltonian satisfies $\sigma_z H(k,z) \sigma_z = -H(k,-z)$, preserving CS with $\Gamma = \sigma_z$.

        \textit{Fundamental constraints from CS.}
        CS imposes algebraic constraints on the Hamiltonian independent of its crystalline and structural origins. For instance, the determinant of Eq.~\eqref{eq:G_chiral_z} gives
        \begin{equation}\label{eq:detcs2}
        \det H(z_0 + z) = (-1)^N \det H(z_0 - z),
        \end{equation}
        where $N$ is the number of bands (or sublattices). At the chiral-symmetric points $z = z_0$, utilizing Floquet periodicity:
        \begin{equation}\label{eq:detcs}
        \det H(z_0) = (-1)^N \det H(z_0).
        \end{equation}
        For odd $N$, this requires $\det H(z_0) = 0$—the Hamiltonian must be singular at chiral-symmetric points.
        
        The trace similarly requires
        \begin{equation}\label{eq:trcs}
        \tr H(z_0 + z) = -\tr H(z_0 - z),
        \end{equation}
        which implies $\tr H(z_0) = 0$ at chiral-symmetric points for any $N$. Physically, this means the on-site potentials must sum to zero: either all vanish individually, or they sum to zero.

        \textit{Relationship between crystalline and CS.}
        These constraints reveal that CS is more fundamental than its crystalline constituents. The logical relationship is $z$-Ref and BpS together imply CS, but CS does not require both—it can arise from the trace and determinant constraints alone.
        This clarifies a point of confusion in the literature: the existence of a $z$-Ref axis alone does \textit{not} guarantee CS in the Floquet operator~\cite{Bellec2017}. Both $z$-Ref and BpS must be present simultaneously, or the trace and determinant conditions must be satisfied. These two cases are demonstrated in Sec.~\ref{subsec:CSEngg} (see Fig.~\ref{fig:1dcs}).

        \subsubsection{$z$-Reversal symmetry ($z$-RS)}
        
    The $z$-RS operator $T_z$ is antiunitary, reversing the propagation direction analogously to time-reversal in quantum mechanics. Writing $ T_z = \mathcal{T}_z \mathcal{K}$, where $\mathcal{T}_z$ is unitary and $\mathcal{K}$ denotes complex conjugation, the symmetry condition reads
    \begin{equation}\label{eq:trs}
    \mathcal{T}_z H(k,z) \mathcal{T}_z^{-1} = H^*(-k,-z).
    \end{equation}
    The squared operator $T_z^2 = \pm 1$ determines the symmetry class: $T_z^2 = +1$ corresponds to spinless (or integer-spin) systems, while $T_z^2 = -1$ corresponds to spinful (half-integer-spin) systems exhibiting Kramers degeneracy. In one-dimensional photonic waveguide arrays, $T_z^2 = +1$ is the relevant case. The $T_z^2 = -1$ scenario, which underlies the quantum spin Hall effect in two dimensions, can be emulated in photonics using polarization as a pseudospin degree of freedom~\cite{Hafezi2011,Maczewsky2020}.

    \textit{Simplification for real coupling coefficients.}
    In photonic waveguide arrays, the evanescent coupling coefficients are generically real~\cite{Szameit2010}.\footnote{Complex effective couplings can arise from synthetic gauge fields~\cite{Wimmer2017}, which enter at the same level as the Bloch quasimomentum $k$.} For a Hamiltonian $H_p$ with real matrix elements in position space, the Bloch Hamiltonian satisfies
    \begin{equation}\label{eq:hp}
    H_p^*(k) = H_p(-k).
    \end{equation}
    That is, complex conjugation combined with $k \to -k$ leaves the Hamiltonian invariant.
    
    Substituting into Eq.~\eqref{eq:trs}, the $z$-reversal condition for photonic systems reduces to
    \begin{equation}\label{eq:trsph}
    \mathcal{T}_z H_p(k,z) \mathcal{T}_z^{-1} = H_p(k, -z),
    \end{equation}

    and for the photonic evolution operator:
    \begin{equation}\label{eq:trspu}
    \mathcal{T}_z U_p(k,z) \mathcal{T}_z^{-1} = U_p^{*}(-k,-z).
    \end{equation}
    For the general (non-photonic) case, $z$-RS constrains the quasienergy dispersion to satisfy $\mathcal{E}(k) = \mathcal{E}(-k)$, with band extrema at the $z$-reversal invariant momenta $k = 0$ and $k = \pi$. For photonic systems with real couplings, this same constraint follows independently from Eq.~\eqref{eq:hp} directly implies $\mathcal{E}_p(k) = \mathcal{E}_p(-k)$. The photonic $z$-RS condition [Eq.~\eqref{eq:trsph}] then provides no additional spectral constraint but instead constrains the $z$-dependence of the driving protocol.

    \textit{Equivalence with $z$-Ref in photonic systems.}
    Eq.~\eqref{eq:trsph} has the same form as the $z$-Ref condition [Eq.~\eqref{eq:zref}]. This is not coincidental: for photonic Hamiltonians with real couplings, $z$-RS effectively reduces to a unitary (rather than antiunitary) operation because complex conjugation acts trivially. Consequently, $z$-RS and $z$-Ref symmetry become equivalent in this context. This equivalence simplifies the symmetry analysis of photonic waveguide arrays. For systems with complex couplings induced by synthetic gauge fields, $z$-RS and $z$-Ref are no longer equivalent. The full antiunitary structure of $\mathcal{T}_z$ must then be retained.

    \subsubsection{Particle-hole symmetry (PHS)}
    The third AZ symmetry, PHS, is represented by the antiunitary operator $C = \mathcal{C} \mathcal{K}$, with $C^2 = +1$ in the systems considered here. In systems possessing both chiral and $z$-RS, we have $\mathcal{C} = \mathcal{T}_z \Gamma$. For the $z$-dependent Bloch Hamiltonian:
    \begin{equation}\label{eq:phs}
    \mathcal{C} H(k,z) \mathcal{C}^{-1} = -H^*(-k,z).
    \end{equation}
    For photonic Hamiltonians with real couplings, this simplifies to
    \begin{equation}\label{eq:phs_photonic}
    \mathcal{C} H_p(k,z) \mathcal{C}^{-1} = -H_p(k,z).
    \end{equation}
    The spectral constraints differ between the general and photonic cases. For the general case, PHS forces quasienergies to appear in pairs $\{\mathcal{E}(k), -\mathcal{E}(-k)\}$: if $|\psi_k\rangle$ is an eigenstate with quasienergy $\mathcal{E}$ at momentum $k$, then $\mathcal{C}|\psi_k\rangle$ is an eigenstate with quasienergy $-\mathcal{E}$ at momentum $-k$. For photonic systems, Eq.~\eqref{eq:phs_photonic} acts at fixed $k$, so PHS directly enforces $\pm\mathcal{E}_p$ pairing at each momentum like CS.
    

    \textbf{\textit{shifted}-PHS ($s$-PHS).}
    In certain systems, PHS holds only about a shifted momentum origin $k_0$:
    \begin{equation}\label{eq:shifted_phs}
    \mathcal{C} H(k + k_0, z) \mathcal{C}^{-1} = -H^*(-k + k_0, z).
    \end{equation}
    We term this \textit{shifted} PHS, because the particle-hole operation is combined with a shift in momentum space. This symmetry, which has been largely overlooked in the literature, arises in non-bipartite lattices and will be examined in detail in Sec.~\ref{sec:engsymph}.

    \textit{Relationship to CS.}
    When BpS is present, the PHS condition [Eq.~\eqref{eq:phs_photonic}] coincides with the CS condition [Eq.~\eqref{eq:css}] for $z$-independent Hamiltonians, and the particle-hole operator can be identified with the sublattice operator as $\mathcal{C} = \Sigma_z$. This identification requires $k_0 = 0$: if the sublattice operator anticommutes with every driving step individually, PHS holds at the standard origin and no momentum shift is needed. For $k_0 \neq 0$, at least one driving step must violate sublattice anticommutation, requiring non-bipartite coupling (see \cite{SM}~\ref{app:3wg_sphs}). In photonic waveguide arrays with real couplings, sublattice-based PHS and $s$-PHS are therefore mutually exclusive: the former requires BpS, the latter exists only in its absence. 


    \textit{Constraints from PHS.}
    The determinant of Eq.~\eqref{eq:phs_photonic} gives
    \begin{equation}\label{eq:detphs}
        \det H_p = (-1)^N \det H_p.
    \end{equation}
    For odd $N$, this requires $\det H_p = 0$—analogous to the CS constraint but holding for all $z$, not just at chiral-symmetric points.
    
    The trace condition reads
    \begin{equation}\label{eq:trphs}
        \tr H_p = 0.
    \end{equation}
    The vanishing trace indicates that diagonal (on-site) terms must cancel across the unit cell. Both the trace and determinant conditions must hold simultaneously for PHS, satisfying one alone is insufficient, as demonstrated in Sec.~\ref{sec:engsymph}. These constraints do not extend to $s$-PHS in the same way since it relates the Hamiltonian at momenta $k+k_0$ and $-k+k_0$, so the trace and determinant conditions become relations between different $k$-points rather than constraints at each $k$ individually.

    \subsubsection{Symmetry compatibility}
    Table~\ref{tab:symmetries} collects the transformation rules and spectral constraints for each symmetry. 
    
    \begin{table*}[!htb]
    \centering
    \begin{tabular}{|c|c|c|c|c|}
    \hline 
    \textbf{Symmetry $(S)$} &  $S H(k ,z)  S^{-1} $ &  $S U(k,z) S^{-1}  $ & $S H_p(k,z) S^{-1} $ & \textbf{Spectral constraint} \\
    \hline
    Bipartite & $-H(k,z)$ & $U^{-1}(k,z)$ & $-H_p(k,z)$ & $\{\mathcal{E}_p(k), -\mathcal{E}_p(k)\} $ \\
    \hline
    $z$-Reflection & $H(k,-z)$ & $U^{-1}(k,-z)$ & $H_p(k,-z)$ & None \\
    \hline
    Chiral & $-H(k,-z)$ & $U(k,-z)$ & $-H_p(k,-z)$ & $\{\mathcal{E}_p(k), -\mathcal{E}_p(k)\} $ \\
    \hline
    $z$-Reversal & $H^*(-k,-z)$ & $U^*(-k,-z)$ & $H_p(k,-z)$ & $\mathcal{E}_p(k) = \mathcal{E}_p(-k)$ \\
    \hline
    Particle-hole & $-H^*(-k,z)$ & $U^*(-k,z)$ & $-H_p(k,z)$ & $\{\mathcal{E}_p(k), -\mathcal{E}_p(k)\} $ \\
    \hline
    \textit{shifted}-PH ($k \rightarrow k + k_0$) & $-H^*(-k+k_0,z)$ & $U^*(-k+k_0,z)$ & $-H_p(-k+k_0,z)$ & $\{\mathcal{E}_p(k+k_0), -\mathcal{E}_p(-k+k_0)\} $ \\
    \hline
    \end{tabular}
    \caption{Symmetry transformations $S$ of the $z$-dependent Hamiltonian $H(k,z)$, evolution operator $U(k,z)$, and photonic Hamiltonian $H_p(k,z)$ (with real couplings), together with the induced constraints on the energy spectrum $\mathcal{E}(k)$. We choose unitary version of $S$ for anti-unitary case.}
    \label{tab:symmetries}
    \end{table*}
    
\begin{figure}[tbh!]
    \centering
    \includegraphics[scale=0.9]{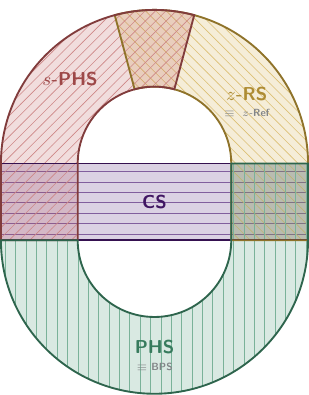}
    \caption{Symmetry relations in $z$-dependent photonic waveguide arrays. Each region represents a fundamental symmetry: CS, $z$-RS, PHS, and $s$-PHS. Overlapping regions indicate cases where multiple symmetries coexist. In the photonic case, the BpS implies both CS and PHS, and $z$-Ref symmetry implies $z$-RS.}
    \label{fig:symgp}
    \end{figure}

    Figure~\ref{fig:symgp} summarizes these compatibility relations. BpS combined with $z$-Ref implies CS; CS combined with $z$-reversal implies PHS and vice versa; for real couplings, $z$-reversal reduces to $z$-Ref. We consider examples for each and overlapping regions in the next section.
    
\section{Engineering symmetries in photonic waveguide arrays}\label{sec:engsymph}

        This section constructs explicit one-dimensional waveguide arrays realizing each of the symmetries introduced above—CS, PHS, \textit{shifted}-PHS, and $z$-RS—and analyzes their topological edge states in finite geometries. We address which symmetries support topologically protected boundary modes. In \cite{SM}~\ref{app:inversion}, we also consider inversion symmetry for the sake of completeness.

    \subsection{Chiral Symmetry}\label{subsec:CSEngg}
    In one-dimensional Floquet systems, CS can be realized in two ways: either by preserving both BpS and \textit{z}-Ref or by breaking both simultaneously while still satisfying the fundamental determinant and trace constraints [Eqs.~\eqref{eq:detcs} and \eqref{eq:trcs}].
    
    \begin{figure}[!htb]
        \centering
        \subfloat[\label{fig:2wg_chiral_symmetry}]{%
            \begin{adjustbox}{valign=b}
                \includegraphics[scale=0.6]{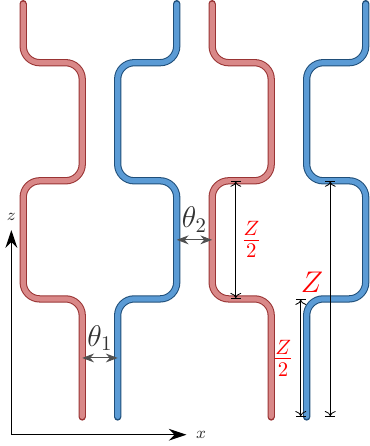}
            \end{adjustbox}%
        }
        \hspace*{0.4cm}
        \subfloat[\label{fig:2wg_chiral_symmetry_breaking}]{%
            \begin{adjustbox}{valign=b, raise=0.2cm}
                \includegraphics[scale=0.61]{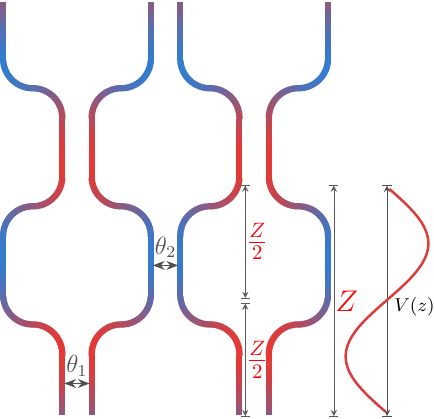}
            \end{adjustbox}%
        }
        
        \caption{In a one dimensional waveguide arrays with two waveguides per unit cell and period \( Z \), there are two cases to preserve CS: (a) zero onsite potential preserves BpS and \textit{z}-Ref, and (b) a sinusoidal time-varying onsite potential $V(z)$ (indicated by color gradient along the \( z \)-axis) breaks both BpS and \textit{z}-Ref.}
        \label{fig:1dcs}
    \end{figure}

    \textit{Case 1: Preserving BpS and $z$-Ref.}
    To engineer CS in the first scenario, one can either implement the structural symmetries (BpS and \textit{z}-Ref) or, equivalently, use the product of PHS and \textit{z}-RS, since $\Gamma = \mathcal{C} \cdot \mathcal{T}_z$. For the two-waveguide network shown in Fig.~\ref{fig:1dcs}(a), the photonic $z-$dependent Hamiltonian is 
    \begin{equation}\label{eq:chiral_photonic_ham1}
    H_p^{(1)}(k) = \begin{pmatrix} 0 & \theta_1 \\ \theta_1 & 0 \end{pmatrix}, \quad 0 \leq z < \frac{Z}{2},
    \end{equation}
    \begin{equation}\label{eq:chiral_photonic_ham2}
    H_p^{(2)}(k) = \begin{pmatrix} 0 & \theta_2 e^{ik} \\ \theta_2 e^{-ik} & 0 \end{pmatrix}, \quad \frac{Z}{2} \leq z < Z.
    \end{equation}
    The vanishing diagonal elements ensure BpS. The given driving steps with proper origin guarantee $z$-Ref symmetry about $z_0 = Z/4$ and $z_0 = 3Z/4$ for each step. A waveguide network similar to Fig.~\ref{fig:1dcs}a has been realized experimentally in Ref.~\cite{Mukherjee2018}. There, the finite propagation length limits the number of accessible driving periods, and the long-time dynamics is recovered by placing the waveguides inside a cavity that recycles the optical state over successive round trips.

    \textit{Case 2: Breaking both structural symmetries.}
    Alternatively, CS can also be achieved by breaking both BpS and \textit{z}-Ref while still satisfying the constraints on determinant and trace [Eqs.~\eqref{eq:detcs}--\eqref{eq:trcs}]. Introducing a $z$-dependent on-site potential $V(z)$ breaks the BpS. If additionally $V(z_0 + z) \neq V(z_0 - z)$ for any $z_0$, then $z$-Ref is also broken. In photonic waveguide arrays, such modulation can be implemented by varying the refractive index along the propagation axis~\cite{Ke2016}.
    
    The resulting Bloch Hamiltonian for the configuration in Fig.~\ref{fig:1dcs}(b) is
    \begin{equation}\label{eq:nonBpS_chiral_photonic_ham1}
    H_p^{(1)}(k,z) = \begin{pmatrix} V\sin(z) & \theta_1 \\ \theta_1 & -V\sin(z) \end{pmatrix}, \quad 0 \leq z < \frac{Z}{2},
    \end{equation}
    \begin{equation}\label{eq:nonBpS_chiral_photonic_ham2}
    H_p^{(2)}(k,z) = \begin{pmatrix} V\sin(z) & \theta_2 e^{ik} \\ \theta_2 e^{-ik} & -V\sin(z) \end{pmatrix}, \quad \frac{Z}{2} \leq z < Z.
    \end{equation}
    This Hamiltonian satisfies $\sigma_z H_p(k,z) \sigma_z = -H_p(k,-z)$, preserving CS with $\Gamma = \sigma_z$, while respecting the constraints Eqs.~\eqref{eq:detcs2}–\eqref{eq:trcs}.
    The opposite-sign diagonal elements satisfy the trace constraint [Eq.~\eqref{eq:trcs}]. The determinant constraint [Eq.~\eqref{eq:detcs}] is likewise satisfied.
    
\begin{figure}[!htb]
        \centering
        \subfloat[\label{fig:geo}]{%
            \includegraphics[scale=0.6]{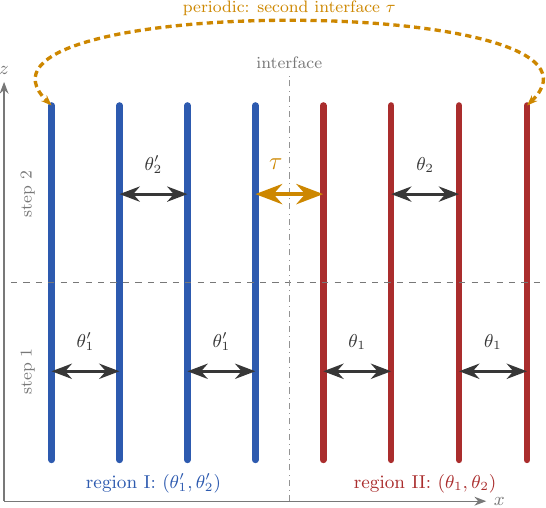}%
        }\hspace*{0.03cm}\\
        \subfloat[\label{fig:2wgcsf}]{%
            \includegraphics[scale=0.215]{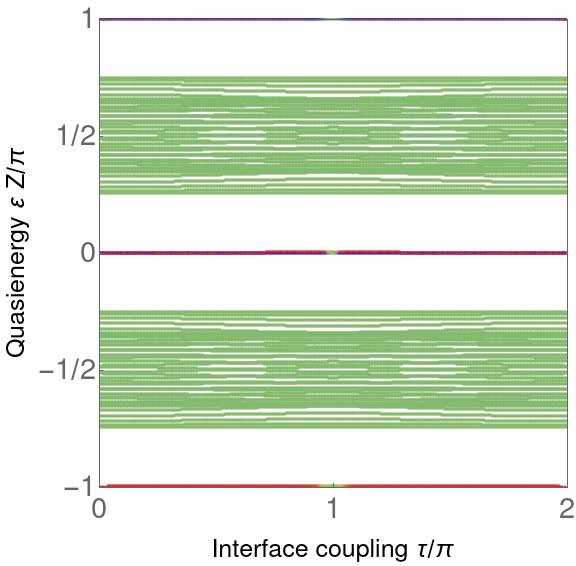}%
        }\hspace*{0.1cm}
        \subfloat[\label{fig:2wg_chiral_breaking}]{%
            \includegraphics[scale=0.186]{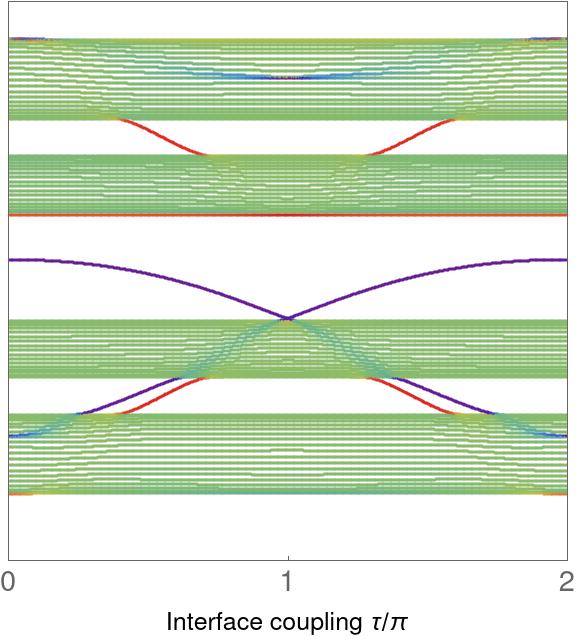}%
        }

        \caption{Topological edge states in a chiral-symmetric Floquet system. (a)~Cylindrical geometry with two regions of distinct topology sharing two interfaces with parameters  $\theta_1 = \pi/2$, $\theta_2 = 3\pi/4$ (red region) and $\theta_1' = \pi/2$, $\theta_2' = \pi/4$ (blue region). (b)~Quasienergy spectrum versus interface coupling $\tau$: edge states at $\varepsilon = 0$ and $\varepsilon = \pi$ remain robust under parameter variation. Green indicates bulk states and red (blue) indicates states localized at the left (right) interface. (c)~Adding a uniform on-site potential of $-0.9$ to $H_p^{(1)}(k)$ breaks CS, coupling edge states to bulk bands and destroying topological protection.}
        \label{fig:1dcsfin}
    \end{figure}

    In one-dimensional Floquet systems, CS protects boundary modes at quasienergies $\varepsilon = 0$ and $\varepsilon = \pi$~\cite{Asboth2014,Fruchart2016,Roy2017}. We analyze a finite geometry of 20 unitcells along the $x$-direction, illustrated in Fig.~\ref{fig:1dcs}a. For numerical convenience, this geometry is coupled at both boundaries to an additional finite network of 20 unitcells with a different set of parameters \(\theta_{j=1,2}\). This configuration forms a cylindrical geometry—periodic along the propagation direction with two domain walls separating topologically distinct regions—that eliminates trivial edge effects arising from open boundaries along $x$ (see Fig.~\ref{fig:1dcsfin}(a)).
    
    In the above geometry, one region satisfies $\theta_1 < \theta_2$ (topologically nontrivial) and the adjacent region satisfies $\theta_1 > \theta_2$ (trivial). Edge states appear at the interface between these two regions at both $\varepsilon = 0$ and $\varepsilon = \pi$, characteristic of chiral-symmetric Floquet systems, as shown in Fig.~\ref{fig:1dcsfin}(b). Varying the interface coupling $\tau$ does not affect the robustness of these edge states. The boundary modes are then counted by a pair of winding numbers $\nu_0$ and $\nu_\pi$ at quasienergy $\varepsilon=0$ and $\varepsilon=\pi$, defined in the two symmetric time frames ~\cite{Asboth2014, Delplace2017}. These invariants apply directly in the photonic setting, as the paraxial propagator over one modulation period is the Floquet evolution operator.
    
    CS can be broken by introducing a symmetric on-site potential at first time step to $H_p^{(1)}(k)$. This couples the edge states to the bulk bands, destroying their topological protection (Fig.~\ref{fig:1dcsfin}(c)). This demonstrates the essential role of CS in protecting zero and $\pi$ modes.

    Systems with CS alone belong to class AIII; those with CS, PHS, and $z$-RS belong to class BDI—both supporting $\mathbb{Z}$ topological invariants in one dimension.

    \subsection{Particle Hole Symmetry}\label{subsec:PHS_engg}

\begin{figure}[!htb]
        \centering
        \subfloat[\label{fig:3wgphs}]{%
            \includegraphics[scale=0.6]{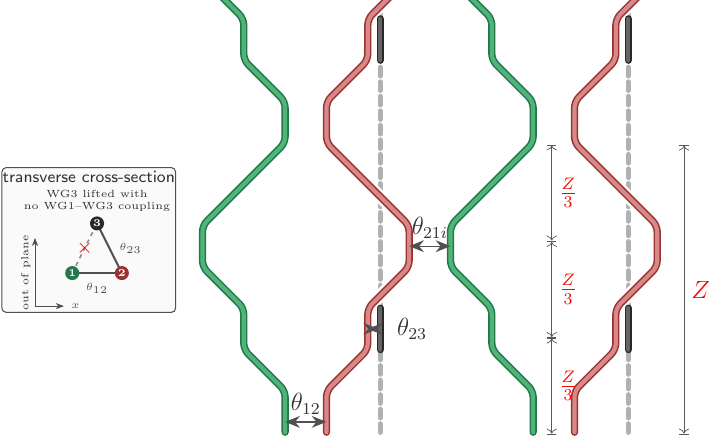}%
        }\\[2mm]
        \subfloat[\label{fig:3wgphsrb}]{%
            \includegraphics[scale=0.215]{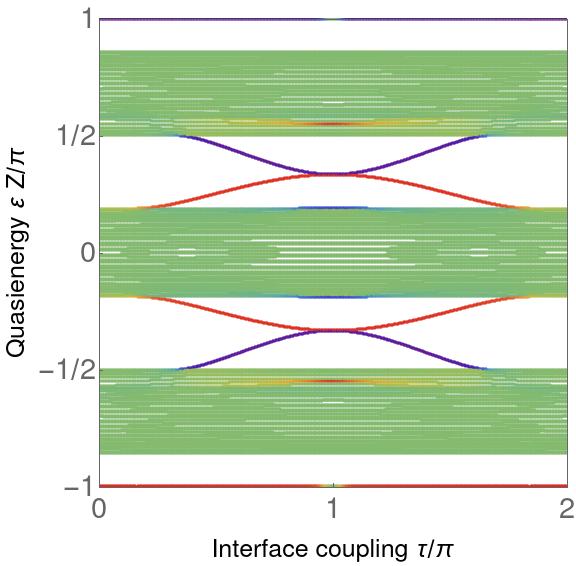}%
        }\hspace*{0.2cm}
        \subfloat[\label{fig:3wgphsbk}]{%
            \includegraphics[scale=0.186]{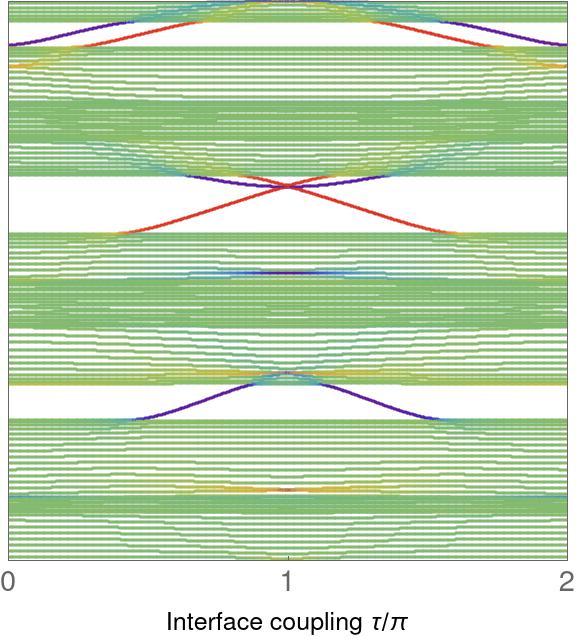}%
        }
    
        \caption{PHS in a three-waveguide bipartite network with parameters $\theta_{12} = 2.4$, $\theta_{23} = 1.9$, $\theta_{21i} = 1.6$, and $\theta_{12}' =\pi/2$, $\theta_{23}' = \pi/2$, $\theta_{21i}' = \pi/4$. (a)~Coupling scheme preserving BpS; the dashed line indicates $\text{WG}_3$ moving out of plane to avoid coupling to $\text{WG}_1$. (b)~Edge states at $\varepsilon = \pi$ remain robust under variation of interface coupling. (c)~Adding an on-site potential of $-0.19$ to all three waveguides during $0 \leq z < Z/3$ breaks BpS (and hence PHS), destroying these topological modes.}
        \label{fig:1dphsfin}
    \end{figure}

    In photonic systems with real coupling coefficients, PHS is equivalent to a BpS, with the sublattice operator $\Sigma_z$ acting as the PHS operator [cf.\ Eq.~\eqref{eq:phs_photonic}]. Although the two-waveguide network in Fig.~\ref{fig:1dcs}(a) already satisfies this symmetry, qualitatively new features arise in systems with an odd number of bands. We therefore consider a three-waveguide (3WG) array with BpS, as shown in Fig.~\ref{fig:1dphsfin}(a).

    The 3WG array is driven using a three-step protocol along the propagation direction $z$, in real space,
    \begin{equation}\label{eq:3wg_coupling}
    H_p(z) =
    \begin{cases}
    \text{WG}_1 \leftrightarrow \text{WG}_2 : \; \theta_{12}, & 0 \leq z < Z/3, \\[1mm]
    \text{WG}_2 \leftrightarrow \text{WG}_3 : \; \theta_{23}, & Z/3 \leq z < 2Z/3, \\[1mm]
    \text{WG}_2 \leftrightarrow \text{WG}_1 : \; \theta_{21i}, & 2Z/3 \leq z < Z.
    \end{cases}
    \end{equation}
    The coupling scheme involves only $\text{WG}_1\leftrightarrow \text{WG}_2$ and $\text{WG}_2\leftrightarrow \text{WG}_3$ interactions, preserving the BpS. As a result, the BpS of the lattice is preserved: $\text{WG}_2$ (shown in red in Fig.~\ref{fig:1dphsfin}(a)) forms one sublattice, while $\text{WG}_1$ and $\text{WG}_3$ belong to the other. The explicit Bloch Hamiltonians and PHS operator form are given in \cite{SM}~\ref{app:hamiltonians}. The resulting geometry may be viewed as a one-dimensional analog of the Lieb lattice~\cite{Weeks2010}.

    For odd $N$, PHS guarantees that bands come in $\{\varepsilon,-\varepsilon\}$ pairs, leaving one unpaired band pinned at $\varepsilon = 0$. This prevents a gap at zero quasienergy. Topological edge states therefore appear only at $\varepsilon = \pi$, as confirmed in Fig.~\ref{fig:1dphsfin}(b). The boundary modes in the $\varepsilon=\pi$ gap is counted by the $\mathbb{Z}_2$ parity $\nu_{\mathrm{PH}}(\pi)$ ~\cite{Nathan2015, RoyHarper2017, Hockendorf2018}.
    
    PHS can be broken by violating the determinant or trace constraints [Eqs.~\eqref{eq:detphs}--\eqref{eq:trphs}]. One approach is to add an on-site potential to all three waveguides during the first time step ($0 \leq z < Z/3$). As shown in Fig.~\ref{fig:1dphsfin}(c), this perturbation destroys the $\varepsilon = \pi$ edge states. Alternatively, introducing direct $\text{WG}_1 \leftrightarrow \text{WG}_3$ coupling breaks BpS. Interestingly, this latter modification does \textit{not} trivialize the topology—instead, a new version of the PHS symmetry emerges to protect the edge states, as discussed next.

    \subsection{\textit{shifted}-Particle Hole Symmetry}\label{subsec:sPHS_engg}
    Restoring the $\text{WG}_1 \leftrightarrow \text{WG}_3$ coupling breaks BpS and conventional PHS, yet edge states persist at $\varepsilon = \pi$. The standard classification predicts trivial topology for one-dimensional systems lacking PHS, CS, and $z$-RS~\cite{Roy2017}. The protecting symmetry is \textit{shifted}-PHS ($s$-PHS), a variant of particle-hole symmetry.

\begin{figure}[!htb]
        \centering
        \subfloat[\label{fig:3wgsphs1}]{%
            \includegraphics[scale=0.5]{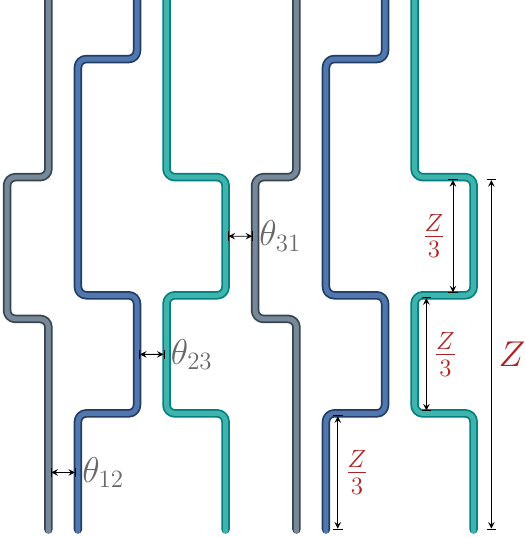}%
        }\\[2mm]
        \subfloat[\label{fig:3wgphsrb1}]{%
            \includegraphics[scale=0.215]{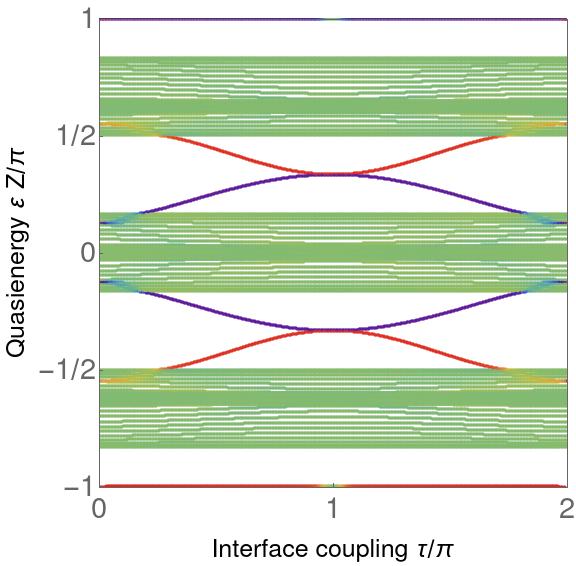}%
        }\hspace*{0.2cm}
        \subfloat[\label{fig:3wgsphsrb2}]{%
            \includegraphics[scale=0.185]{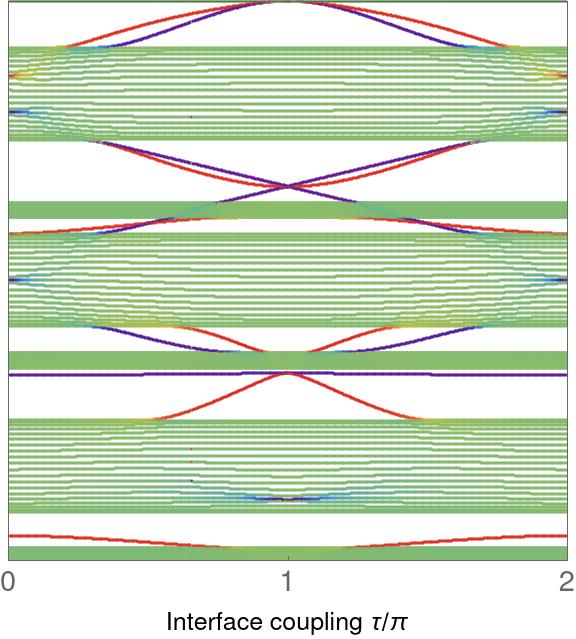}%
        }
    
        \caption{$s$-PHS in a 3WGs network with cyclic coupling with parameters $\theta_{12} = \pi/2 - 0.3, \theta_{23} = \pi/2 + 0.9, \theta_{31} = \pi - 0.2,$ and $\theta'_{12} = \pi/2, \theta'_{23} = \pi/2, \theta'_{31} = \pi/4.$ (a)~Network geometry with cyclic coupling, which breaks the BpS. (b)~In the absence of conventional particle--hole symmetry, edge states nevertheless occur at quasienergy $\varepsilon = \pi$. (c)~Adding an on-site potential of $0.9$ to all three waveguides during $0 \leq z < Z/3$ breaks $s$-PHS, destroying these topological modes.}
        \label{fig:1dsphsfin}
    \end{figure}

    The 3WGs network with cyclic coupling (Fig.~\ref{fig:1dsphsfin}(a)) realizes $s$-PHS with
    \begin{equation}\label{eq:sphs_coupling}
    H_p(z) = \begin{cases}
    \text{WG}_1 \leftrightarrow \text{WG}_2: \; \theta_{12}, & 0 \leq z < Z/3, \\[1mm]
    \text{WG}_2 \leftrightarrow \text{WG}_3: \; \theta_{23}, & Z/3 \leq z < 2Z/3, \\[1mm]
    \text{WG}_3 \leftrightarrow \text{WG}_1: \; \theta_{31}, & 2Z/3 \leq z < Z.
    \end{cases}
    \end{equation}
    The all-to-all coupling pattern breaks the BpS (all three waveguides now couple to each other), and the asymmetric driving breaks $z$-Ref symmetry. Consequently, CS, PHS, and $z$-RS are all absent.
    
    In the absence of conventional protecting symmetries, edge states are nevertheless observed at quasienergy $\varepsilon = \pi$ [Fig.~\ref{fig:1dsphsfin}(b)]. The protecting symmetry is $s$-PHS. Under a momentum shift $k \rightarrow k + k_0$, the quasienergy spectrum appears in particle-hole related pairs $\{\varepsilon(k + k_0), -\varepsilon(-k + k_0)\}$. For the present network, the shifted momentum $k_0$  takes the value $ \pm \pi/2$ (see \cite{SM}~\ref{app:3wg_sphs} for details). The odd number of sites per unit cell forbids the Pfaffian invariant of the even-band case and pins an unremovable zero-quasi-energy mode, which leaves the $\pi$ gap as the only host of protected boundary states. The parity $\nu_\pi$ of Eq.~\eqref{eq:nuPi} is the $\pi$-gap $\mathbb{Z}_2$ invariant of the Floquet particle-hole classification~\cite{Nathan2015}, here carried by the shifted momenta $k_I=\pi/2,\,3\pi/2 $ (see \cite{SM} C for more details on invariant and phase diagram).
    
    Breaking $s$-PHS confirms its role: a constant on-site potential applied to all three waveguides during first step of evolution $0 \leq z < Z/3$ breaks the $s$-PHS constraint and removes the $\varepsilon = \pi$ edge states [Fig.~\ref{fig:1dsphsfin}(c)].

    \subsection{$z$-Reversal Symmetry}\label{subsec:zRS_engg}

    To isolate the role of $z$-RS, we design a network that preserves $z$-RS while breaking both CS and PHS (also $s$-PHS). This can be achieved with a three-waveguide unit cell featuring cyclic coupling within each half-period.
    
\begin{figure}[!htb]
        \centering
        \subfloat[\label{fig:3wgzrs}]{%
            \includegraphics[scale=0.55]{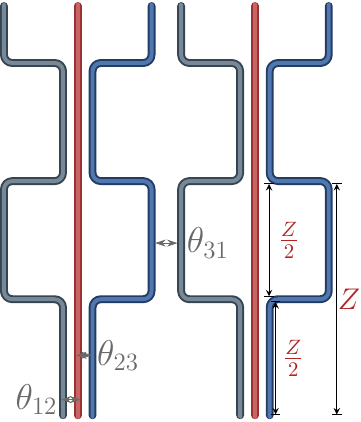}%
        }
        \subfloat[\label{fig:3wgzrsf}]{%
            \includegraphics[scale=0.195]{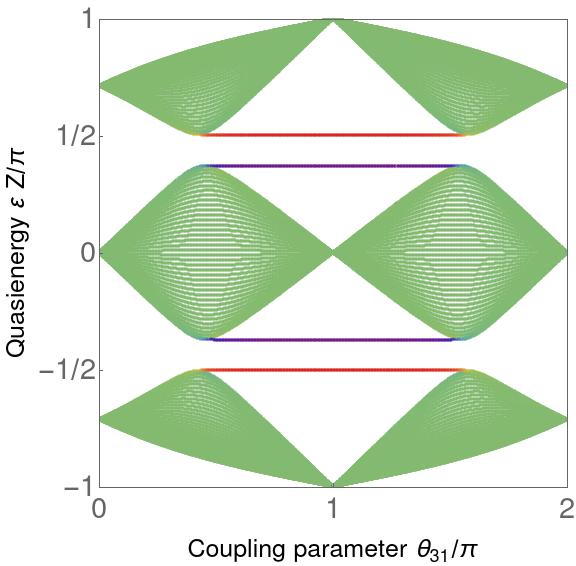}%
        }

        \caption{$z$-RS in a three-waveguide network with non-bipartite cyclic coupling, using parameters $\theta_{12} = \pi/2 - 0.2$, $\theta_{23} = \pi/2 + 0.2$, and $\theta_{31} = \pi/2 - 0.2$. (a)~Three-waveguide network with $z$-Ref-symmetric driving and a non-bipartite coupling structure. (b)~Quasienergy spectrum of a 40-unit-cell system with open boundaries, showing the absence of edge states.}
        \label{fig:3wg_zrs_network}
    \end{figure}

    The configuration in Fig.~\ref{fig:3wg_zrs_network}(a), in real space, employs
    \begin{equation}\label{eq:zrs_coupling}
        H_p(z) =
        \begin{cases}
        \begin{rcases*}
        \text{WG}_1 \leftrightarrow \text{WG}_2 : \theta_{12}, \\
        \text{WG}_2 \leftrightarrow \text{WG}_3 : \theta_{23},
        \end{rcases*}
        & 0 \leq z < Z/2, \\[2mm]
        \text{WG}_3 \leftrightarrow \text{WG}_1 : \theta_{31},
        & Z/2 \leq z < Z .
        \end{cases}
    \end{equation}
    In the first half-period, couplings $\theta_{12}$ and $\theta_{23}$ are present simultaneously; in the second half-period, only the inter-cell coupling $\theta_{31}$ is active.
    
    The symmetric driving protocol, enforces $z$-Ref symmetry, which for real coupling coefficients reduces to $z$-RS. The inter-cell coupling $\text{WG}_3 \leftrightarrow \text{WG}_1'$ breaks the BpS and removes both CS and PHS.
    
    In one dimension, $z$-RS alone (class AI in the AZ classification) does not support nontrivial topology—the classification is trivial~\cite{Schnyder2008,Chiu2016}. This is confirmed by analyzing a 40-unit-cell array with open boundaries.\footnote{The cylindrical geometry used earlier requires two topologically distinct regions. Here with the non-existence of a nontrivial phase, we use open boundaries instead.} Regardless of the coupling parameters, only trivial boundary-dependent states appear (Fig.~\ref{fig:3wg_zrs_network}(b)), confirming that $z$-RS alone is insufficient for topological protection in one dimension. Inversion symmetry likewise yields no protected edge states in one dimension
(\cite{SM}~\ref{app:inversion}).

\section{Conclusion}\label{sec:conc}
In one-dimensional photonic waveguide arrays with periodic $z$-modulation, discrete symmetries determine which boundary states are topologically protected. These symmetries arise from the lattice connectivity and the driving protocol, linking structural properties directly to the AZ classification. For systems possessing chiral, $z$-reversal, and particle-hole symmetry, the classification is class BDI, which supports protected edge states at quasienergies $\varepsilon = 0$ and $\varepsilon = \pi$.

CS arises when both bipartite structure and $z$-reflection symmetry are present, or—in their absence—when the Hamiltonian satisfies the trace and determinant constraints of Eqs.~\eqref{eq:trcs} and \eqref{eq:detcs}. For real coupling coefficients, $z$-reflection implies $z$-reversal symmmtery, and bipartite structure implies particle-hole symmetry. In systems with an odd number of bands, particle-hole symmetry pins one band at $\varepsilon = 0$, precluding a bulk gap there; boundary states therefore occur only at $\varepsilon = \pi$.

Beyond the standard classification, we identified \textit{shifted}-particle-hole symmetry—a particle-hole symmetry centered at non-zero momentum $k_0$—which stabilizes $\varepsilon = \pi$ edge states in non-bipartite networks lacking CS, conventional particle-hole symmetry, and $z$-RS. Standard AZ classification predicts trivial topology for such systems; $s$-PHS provides protection outside this framework and has been largely overlooked in the literature. We demonstrated this symmetry using a three-waveguide cyclic network as a minimal example. Importantly, conventional PHS and $s$-PHS are mutually exclusive: the former requires bipartite structure, while the latter exists only in its absence.

The waveguide networks presented here are readily realizable using femtosecond laser writing techniques~\cite{Szameit2010}, offering a direct experimental platform for probing both standard and \textit{shifted}-particle-hole symmetry-protected topological phases. Moreover, the identification of $s$-PHS expands the landscape of symmetry-protected topological phases in photonic systems and suggests that analogous shifted symmetries may play a role in other driven systems beyond the photonic platform. The symmetry relations along with the constraints derived here—Eqs.\eqref{eq:detcs2}–\eqref{eq:trcs},\eqref{eq:detphs}-\eqref{eq:trphs}—are dimension-independent, so the correspondence between structural properties and AZ classes extends directly to higher-dimensional waveguide arrays.

\begin{acknowledgments}
We would like to thank Pierre~Delplace for very valuable and insightful discussions that helped in many aspects. This research was supported by the Marie Skłodowska-Curie Actions Postdoctoral Fellowship (MSCA-PF) with Project number CayLat-101283743 under the Horizon Europe program and the Starting Grant No. 211310 by the Swiss National Science Foundation.
\end{acknowledgments}

\bibliography{references}
     
\newpage
\clearpage
\onecolumngrid
\appendix
\renewcommand{\thefigure}{A\arabic{figure}}
\setcounter{figure}{0}
\title{Engineering symmetry-protected topological states in waveguide arrays}
\author{Lavi K. Upreti}
\affiliation{Department of Physics, University of Z\"urich, Switzerland}
\maketitle

\section{Hamiltonian description}\label{app:hamiltonians}
The main text specifies each waveguide network through its real-space coupling
scheme. Here we give the corresponding photonic Bloch Hamiltonians $H_p(k,z)$
explicitly. The Floquet evolution operator for each network follows from
$U_F = T\prod_j e^{-i H_p^{(j)} \Delta z_j}$, where $\Delta z_j$ is the
duration of the $j$-th driving step and $T$ is the time-ordering.

The two-waveguide Bloch Hamiltonians for CS with bipartite structure and
$z$-reflection are given in
Eqs.~\eqref{eq:chiral_photonic_ham1}--\eqref{eq:chiral_photonic_ham2},
and without both in
Eqs.~\eqref{eq:nonBpS_chiral_photonic_ham1}--\eqref{eq:nonBpS_chiral_photonic_ham2}.

    \subsection{Three-waveguide network for particle-hole 
    symmetry}\label{app:3wg_cyc}
    The coupling scheme in Eq.~\eqref{eq:3wg_coupling} yields a three-step Bloch Hamiltonian in the basis $(\text{WG}_1, \text{WG}_2, \text{WG}_3)$:
    \begin{equation}
    H_p^{(1)}(k) = \begin{pmatrix} 0 & \theta_{12} & 0 \\ \theta_{12} & 0 & 0 \\ 0 & 0 & 0 \end{pmatrix}, \quad 0 \leq z < Z/3,
    \end{equation}
    \begin{equation}
    H_p^{(2)}(k) = \begin{pmatrix} 0 & 0 & 0 \\ 0 & 0 & \theta_{23} \\ 0 & \theta_{23} & 0 \end{pmatrix}, \quad Z/3 \leq z < 2Z/3,
    \end{equation}
    \begin{equation}
    H_p^{(3)}(k) = \begin{pmatrix} 0 & \theta_{21i}\, e^{ik} & 0 \\ \theta_{21i}\, e^{-ik} & 0 & 0 \\ 0 & 0 & 0 \end{pmatrix}, \quad 2Z/3 \leq z < Z.
    \end{equation}
    The coupling scheme involves only $\text{WG}_1 \leftrightarrow 
    \text{WG}_2$ and $\text{WG}_2 \leftrightarrow \text{WG}_3$ interactions, preserving bipartite structure with $\text{WG}_2$ as one sublattice and $\text{WG}_1$, $\text{WG}_3$ as the other. The sublattice operator 
    $\Sigma_z = \mathrm{diag}(1,-1,1)$ anticommutes with each $H_p^{(j)}(k)$ and serves as the PHS operator at $k_0 = 0$.  
    

    \subsection{Three-waveguide cyclic network and shifted-particle-hole 
    symmetry}\label{app:3wg_sphs}
    
    The cyclic coupling scheme differs from the bipartite network only in the third driving step, where the inter-cell coupling connects $\text{WG}_3$ to $\text{WG}_1$:
    \begin{equation}
    H_p^{(1)}(k) = \begin{pmatrix} 0 & \theta_{12} & 0 \\
    \theta_{12} & 0 & 0 \\ 0 & 0 & 0 \end{pmatrix}, 
    \quad 0 \leq z < Z/3,
    \end{equation}
    \begin{equation}
    H_p^{(2)}(k) = \begin{pmatrix} 0 & 0 & 0 \\
    0 & 0 & \theta_{23} \\ 0 & \theta_{23} & 0 \end{pmatrix},
    \quad Z/3 \leq z < 2Z/3,
    \end{equation}
    \begin{equation}
    H_p^{(3)}(k) = \begin{pmatrix} 0 & 0 & \theta_{31}\,e^{ik} \\
    0 & 0 & 0 \\ \theta_{31}\,e^{-ik} & 0 & 0 \end{pmatrix},
    \quad 2Z/3 \leq z < Z.
    \label{eq:cyclic_H3}
    \end{equation}
    $H_p^{(3)}$ couples $\text{WG}_1$ and $\text{WG}_3$ directly, two sites that carry the same sublattice eigenvalue $+1$ under $\Sigma_z = \mathrm{diag}(1,-1,1)$. The system is therefore no longer bipartite, and a direct calculation gives
    \begin{equation}
    \{\Sigma_z,\, H_p^{(3)}(k)\} 
    = 2\theta_{31}
    \begin{pmatrix} 0 & 0 & e^{ik} \\ 0 & 0 & 0 \\ e^{-ik} & 0 & 0
    \end{pmatrix} \neq 0,
    \end{equation}
    so standard PHS at $k_0 = 0$ is absent. The shifted condition [Eq.~\eqref{eq:sphs_condition_app}] with $k_0 = \pi/2$ is satisfied: $H_p^{(1)}$ and $H_p^{(2)}$ are $k$-independent and anticommute with $\Sigma_z$ by construction, while for $H_p^{(3)}$ a direct substitution gives
    \begin{equation}
    \Sigma_z\, H_p(k + k_0, z)\, \Sigma_z^{-1} 
    = -H_p^*(-k + k_0, z),
    \label{eq:sphs_condition_app}
    \end{equation}
    %
    
    For our case where $k_0 = \pi/2$, $H_p^{(1)}$ and $H_p^{(2)}$, which are 
    $k$-independent, this reduces to the standard anticommutation relation, 
    which holds by construction. For $H_p^{(3)}$, substituting 
    $k \to k + \pi/2$ and $k \to -k + \pi/2$ gives
    \begin{equation}
    \Sigma_z\, H_p^{(3)}\!\left(k+\tfrac{\pi}{2}\right)\Sigma_z^{-1}
    = \theta_{31}
    \begin{pmatrix} 0 & 0 & ie^{ik} \\ 0 & 0 & 0 \\ -ie^{-ik} & 0 & 0
    \end{pmatrix},
    \end{equation}
    \begin{equation}
    -\left[H_p^{(3)}\!\left(-k+\tfrac{\pi}{2}\right)\right]^*
    = \theta_{31}
    \begin{pmatrix} 0 & 0 & ie^{ik} \\ 0 & 0 & 0 \\ -ie^{-ik} & 0 & 0
    \end{pmatrix},
    \end{equation}
    so Eq.~\eqref{eq:sphs_condition_app} is satisfied at $k_0 = \pi/2$. 
    The operator $\Sigma_z$ thus serves as the $s$-PHS operator for the 
    cyclic network. 

\section{Inversion symmetry}\label{app:inversion}
    
    Inversion symmetry acts on the Bloch Hamiltonian as
    \begin{equation}
        \mathcal{P} H(k,z)\, \mathcal{P}^{-1} = H(-k, z).
    \end{equation}
    In contrast to chiral and particle-hole symmetries, inversion symmetry alone does not enforce a symmetric quasienergy spectrum. In one dimension, it yields a $\mathbb{Z}_2$ classification only in the presence of additional symmetries.
    
\begin{figure}[!htb]
        \centering
        \subfloat[\label{sfig:2wg_invs}]{%
            \includegraphics[scale=0.43]{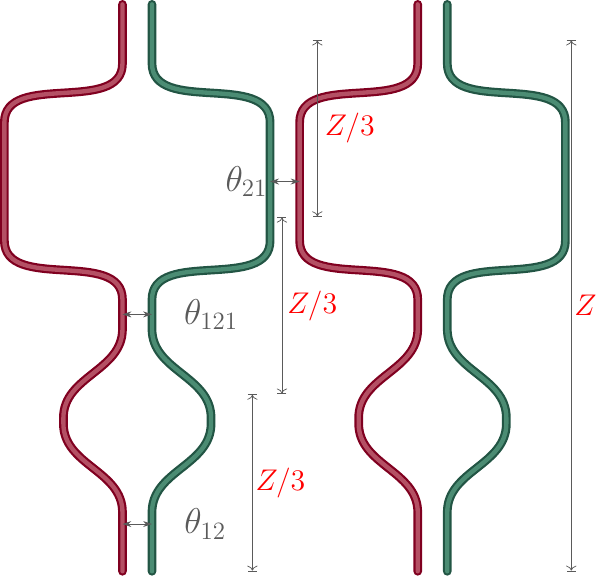}%
        }
        \subfloat[\label{sfig:2wg_invs_finite}]{%
            \includegraphics[scale=0.205]{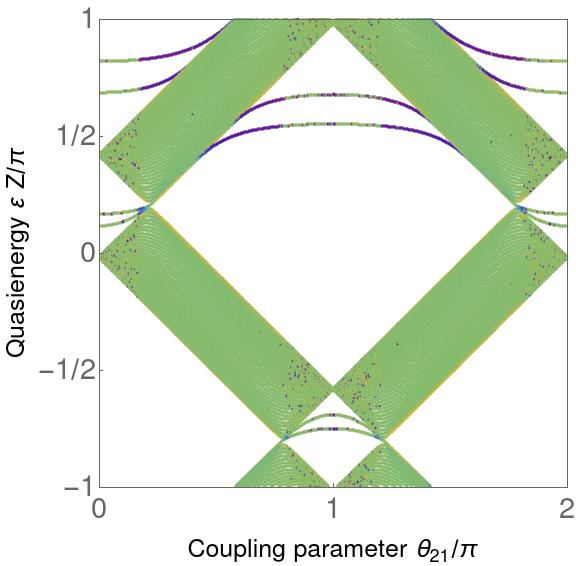}%
        }
        
        \caption{Inversion symmetry without topological protection. (a)~Two-waveguide network preserving inversion symmetry but breaking $z$-reflection and particle-hole symmetry. (b)~Quasienergy spectrum shows only trivial boundary-dependent states, confirming that inversion symmetry alone does not induce nontrivial topology in one dimension.}
        \label{fig:2wg_invs}
    \end{figure}

    The two-waveguide configuration shown in Fig.~\ref{fig:2wg_invs}\subref{sfig:2wg_invs} realizes inversion symmetry via a three-step driving protocol with two distinct intra-cell couplings:
   \begin{equation}\label{eq:inv_coupling}
        H_p(z) = \begin{cases}
                    \text{WG}_1 \leftrightarrow \text{WG}_2: \; \theta_{12}, & 0 \leq z < Z/3, \\[1mm]
                    \text{WG}_1 \leftrightarrow \text{WG}_2: \; \theta_{121}, & Z/3 \leq z < 2Z/3, \\[1mm]
                    \text{WG}_2 \leftrightarrow \text{WG}_1: \; \theta_{21}, & 2Z/3 \leq z < Z.
                \end{cases}
    \end{equation}
    The use of two different intra-cell couplings ($\theta_{12} \neq \theta_{12}'$) breaks $z$-reflection symmetry while preserving spatial inversion. An identical constant on-site potential is added to both waveguides to break PHS (the bipartite trace condition) while maintaining inversion.
    
    \textit{Trivial topology.}
    The finite system supports only termination-dependent edge features, with no boundary modes protected by bulk topology, as shown in Fig.~\ref{fig:2wg_invs}\subref{sfig:2wg_invs_finite}. In the absence of chiral or particle-hole symmetry, inversion symmetry alone does not protect edge states in one-dimensional Floquet systems.

 \section{Topological invariant for the shifted particle-hole symmetry}
\label{app:invariant}
The boundary states protected by the shifted particle-hole symmetry occur at quasi-energy
$\varepsilon=\pi$, a feature absent from static systems, and the topological invariant of the
one-period effective Hamiltonian leaves them undetected. The characterization rests instead on
the evolution operator $U(k,z)$ throughout the driving cycle $0\le z\le Z$, through its phase
bands~Ref[43] (main text). The phase bands $\phi_n(k,z)$ follow from
$U(k,z)\,|\psi_n(k,z)\rangle=e^{-i\phi_n(k,z)}\,|\psi_n(k,z)\rangle$, originate at
$\phi_n(k,0)=0$, and reach the quasi-energies $\varepsilon_n(k)$ at $z=Z$.

\begin{figure}[t]
  \centering
  \includegraphics[width=0.35\columnwidth]{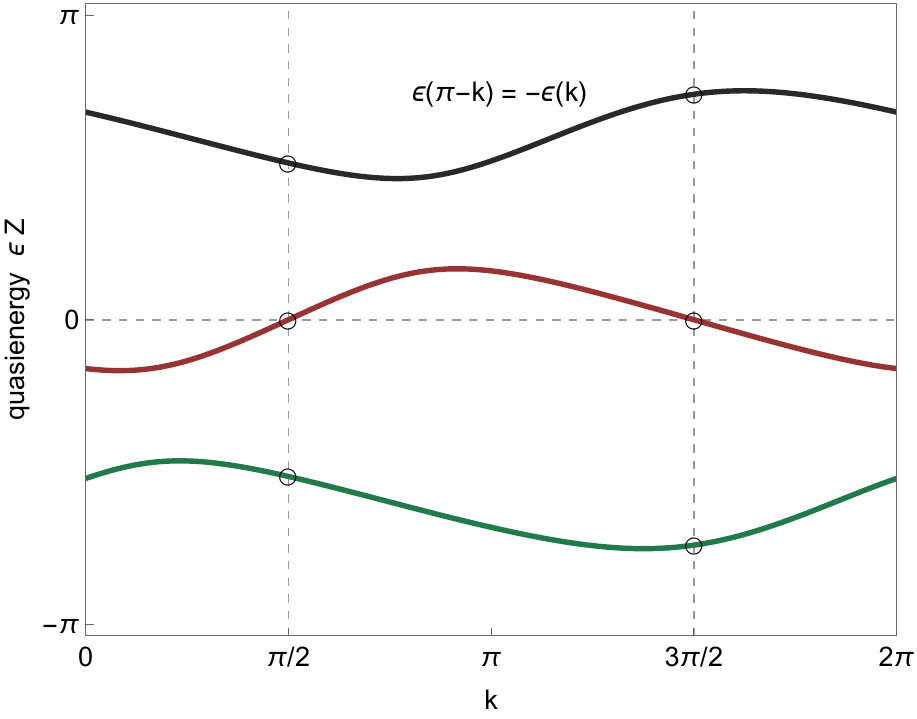}
  \caption{Bulk quasienergy bands of the shifted-particle-hole-symmetric
  network with parameters $\theta_{12}=2.4$, $\theta_{23}=1.9$,
  $\theta_{21i}=1.6$. The spectrum obeys $\varepsilon(\pi-k)=-\varepsilon(k)$,
  the defining relation of the shifted particle-hole symmetry. At the
  self-conjugate momenta $k=\pi/2$ and $3\pi/2$ (dashed) one band is pinned
  at $\varepsilon=0$ and the remaining two form the conjugate pair
  $\pm\varphi$ (open circles).}
  \label{fig:sphs_bands}
\end{figure}

The shifted particle-hole symmetry constrains the Bloch Hamiltonian as
\begin{equation}
C\,H(k,z)\,C^{-1}=-H(\pi-k,z),\qquad C=\mathcal{C}\,\mathcal{K},
\label{eq:sphs}
\end{equation}
with $\mathcal{C}=\mathrm{diag}(1,-1,1)$ in the waveguide
basis (main text). The momentum shift $k\to\pi-k$ sets Eq.~\eqref{eq:sphs} apart from conventional
particle-hole symmetry, where the conjugate momentum is $-k$. Note that we have absorbed $k_0$ by redefining $k \rightarrow k - k_0$ in Eq.\eqref{eq:sphs}.
The same relation holds for the evolution operator,
\begin{equation}
C\,U(k,z)\,C^{-1}=U(\pi-k,z),
\label{eq:sphsU}
\end{equation}
and pairs the phase bands of conjugate partners across the shifted momenta,
\begin{equation}
\phi_{\bar n}(k,z)=-\phi_n(\pi-k,z),
\label{eq:pairing}
\end{equation}
where $n$ is a discrete index.
The bulk quasienergy bands, shown in Fig.~\ref{fig:sphs_bands}, make the
symmetry explicit: the spectrum satisfies $\varepsilon(\pi-k)=-\varepsilon(k)$,
and at the self-conjugate momenta $k=\pi/2$ and $3\pi/2$ one band is pinned at
$\varepsilon=0$ while the other two form a conjugate pair.

The momenta that Eq.~\eqref{eq:sphs} leaves invariant satisfy $\pi-k_I=k_I$ (also evident from the Fig.~\ref{fig:sphs_bands}), hence
\begin{equation}
k_I=\frac{\pi}{2},\quad\frac{3\pi}{2}.
\label{eq:kstar}
\end{equation}
These two points replace $k=0,\pi$ of the unshifted case and carry the topological content. At these invariant momenta
$k_I$ the pairing of Eq.~\eqref{eq:pairing} relates two bands at a single momentum,
$\phi_{\bar n}(k_I,z)=-\phi_n(k_I,z)$, and a conjugate pair becomes degenerate only where the
two phases coincide modulo $2\pi$,
\begin{equation}
\phi_n=-\phi_n\pmod{2\pi}\ \Longrightarrow\ \phi_n\in\{0,\pi\}.
\label{eq:meet}
\end{equation}
A degeneracy at $\phi=\pi$ inside the cycle is topologically protected. Equation~\eqref{eq:sphsU} makes
$U(k_I,z)$ real in the symmetry basis, so the two touching bands span a subspace that evolves as a real rotation,
\begin{equation}
M(z)=\pm\,e^{-i\lambda(z-z_0)\sigma_y},
\label{eq:Mform}
\end{equation}
with $\lambda$ real and the sign fixing the crossing at $\phi=\pi$ for $-$ and at $\phi=0$ for
$+$. The slope $\lambda$ and the location $z_0$ are the only parameters in $M(z)$, and neither lifts
the degeneracy. A single phase-$\pi$ crossing is therefore protected, and crossings appear or vanish
only in pairs.

\begin{figure}[t]
  \centering
  \includegraphics[width=0.49\linewidth]{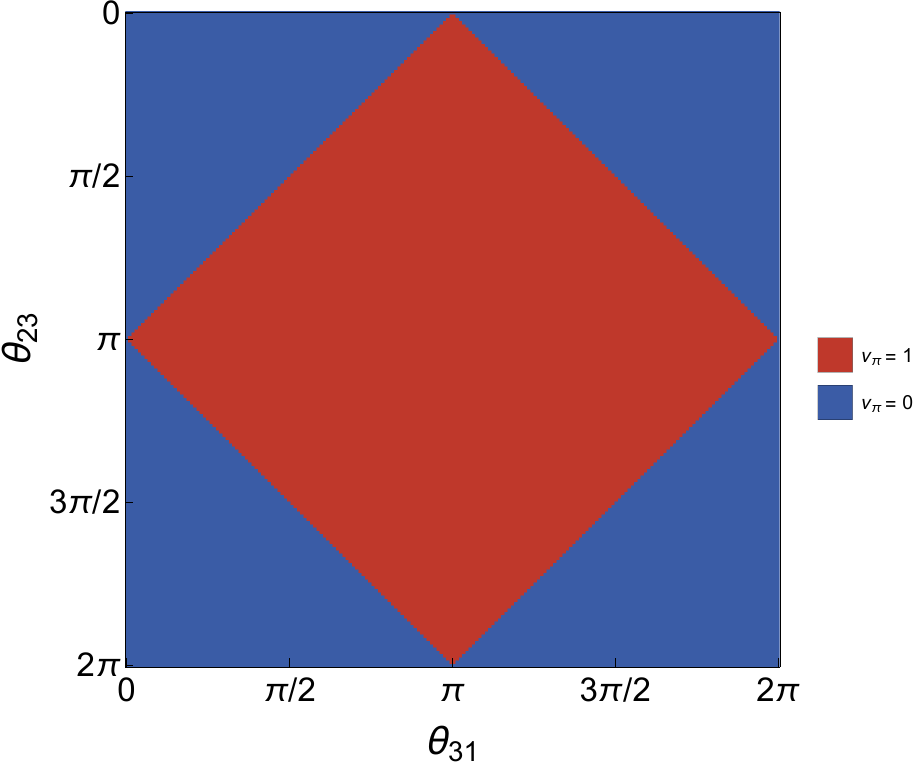}\hfill
  \includegraphics[width=0.49\linewidth]{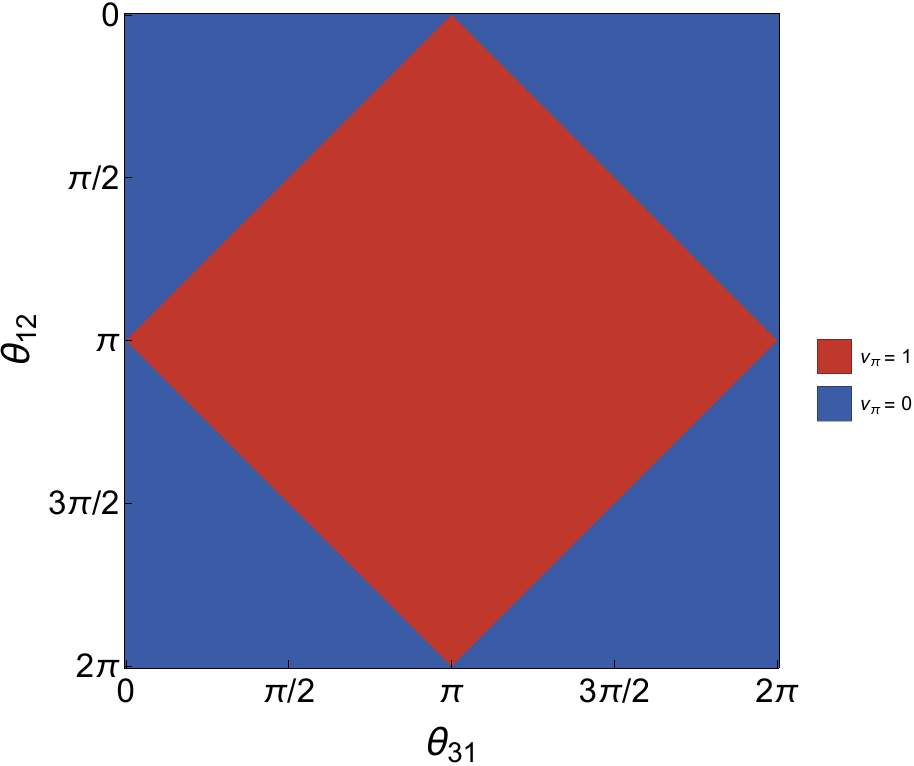}
  \caption{Phase diagram for shifted particle-hole for (left) $(\theta_{23},\theta_{31})$ plane at
  $\theta_{12}=3\pi/2$, and (right) $(\theta_{12},\theta_{31})$ plane at
  $\theta_{23}=\pi/2$. The plot shows the invariant $\nu_\pi$ of Eq.~\eqref{eq:nuPi}, where blue (red) region shows the trivial (topological) regime.}
  \label{fig:phasediagram}
\end{figure}

The parity of their number is the invariant. With $Q_{k_I}$ the number of phase-$\pi$ crossings
at $k_I$ over $0<z<Z$, the count of $\pi$-gap boundary states on each edge equals
\begin{equation}
\nu_\pi=\big(Q_{\pi/2}+Q_{3\pi/2}\big)\bmod 2.
\label{eq:nuPi}
\end{equation}
A real-space shift $k\to k+\pi$ exchanges the two invariant momenta and leaves the edge spectrum
intact, so they enter symmetrically. Each crossing marks a value of $z$ where an eigenvalue of
$U(k_I,z)$ reaches $-1$. The distance from the nearest eigenvalue to $-1$,
\begin{equation}
g(z)=\min_n\big|\,\lambda_n\!\big(U(k_I,z)\big)+1\,\big|,
\label{eq:gz}
\end{equation}
vanishes at every crossing. The traceless drive fixes $\det U(k_I,z)=1$, so the real $U(k_I,z)$ is a rotation with spectrum $\{1,e^{\pm i\phi}\}$ — one band held at $\phi=0$ across the cycle, the conjugate pair carrying every crossing. The distance reduces to $g(z)=\sqrt{1+\operatorname{Tr}U(k_I,z)}$, and $\operatorname{Tr}U(k_I,z)=-1$ marks each phase-$\pi$ crossing. The odd number of sites per unit cell forbids a Pfaffian expression and pins an unremovable zero-quasi-energy mode (Fig 8 b-c of the main text), which leaves the $\pi$ gap as the sole host of protected boundary states. The phase diagram is shown in Fig.~\ref{fig:phasediagram} (left) that maps $\nu_\pi$ across the $(\theta_{23},\theta_{31})$ plane at $\theta_{12}=3\pi/2$ and Fig.~\ref{fig:phasediagram} (right) maps $\nu_\pi$ across the $(\theta_{12},\theta_{31})$ plane at $\theta_{23}=\pi/2$.

\end{document}